\documentclass[a4paper,11pt]{article}
\usepackage{jinstpub} % for details on the use of the package, please see the JINST-author-manual
\newcommand{\dCP}{\ensuremath{\delta_{\mathrm{CP}}}}
\newcommand{\dCPn}{\ensuremath{\delta_{\mathrm{CP}}}^{0}}
\newcommand{\dCPt}{\ensuremath{\delta_{\mathrm{CP}}}^{\mathrm{test}}}
\newcommand{\nue}{\ensuremath{\nu_e}}
\newcommand{\nueb}{\ensuremath{\bar{\nu}_e}}
\newcommand{\numu}{\ensuremath{\nu_{\mu}}}
\newcommand{\numub}{\ensuremath{\bar{\nu}_{\mu}}}

\newcommand\brabar{\textbf{\scalebox{.35}{(}\raisebox{-1.2pt}{--}\scalebox{.35}{)}}}
\makeatletter
\newcommand{\oset}[2]{%
  {\mathop{#2}\limits^{\vbox to -.5\ex@{\kern-\tw@\ex@
   \hbox{\scriptsize #1}\vss}}}}
\makeatother
\newcommand{\nua}[1]{\oset{\brabar}{\nu}_{\!\!#1}}

\title{\boldmath The sensitivity of liquid scintillator detectors to CP-violation with atmospheric neutrinos}

\author[1]{T. Birkenfeld\note{Corresponding author.}}
\author{and A. Stahl}
\affiliation{III. Physics Institute B, RWTH Aachen University,\\Sommerfeldstraße 16, 52066 Aachen, Germany}

\emailAdd{thilo.birkenfeld@rwth-aachen.de}

\abstract{The detection of CP violation in neutrino oscillations is one of the most important goals of the next generation of neutrino experiments.
Here we study the detectability of the CP-violating phase in the oscillation of atmospheric neutrinos.
Liquid scintillator detectors of a few kilotons can probe the low-energy range of the atmospheric neutrino flux. 
We calculate the expected rate, spectrum, and zenith angle distribution for a typical liquid scintillator detector for different detector sites. 
We include a typical detector response with different capabilities for flavour identification and a background model. 
The sensitivity is estimated using a Poisson likelihood analysis. 
}


\arxivnumber{2507.07598} % Only if you have one

\begin{document}
\maketitle
\flushbottom

\section{Introduction}
\label{sec:intro}

The Pontecorvo-Maki-Nakagawa-Sakata matrix describing neutrino oscillations, contains a single CP-violating phase \dCP{}~\cite{Pontecorvo:1967fh, Maki1962}, which can not be removed by regauging of the Dirac fields and can be observed in neutrino oscillations.
It introduces a difference in the flavour conversion probabilities for neutrinos and antineutrinos.
It might be related to leptogenesis~\cite{Fukugita:1986hr, BUCHMULLER199673, BUCHMULLER2005305, BUCHMULLER2005, Pilaftsis1997, Pascoli2007, Hagedorn:2018, Branco2012} in the early universe and the matter-antimatter asymmetry.
For the determination of \dCP{} a setup is needed that measures the different oscillations of the neutrinos and the antineutrinos.
First attempts to measure \dCP{} with neutrino/antineutrino beams and from atmospheric neutrinos have been attempted~\cite{T2K:2023smv, NOvA:2021nfi, Super-Kamiokande:2017yvm} in recent years.
The results were introduced in global oscillation parameter fits~\cite{deSalas:2020pgw, Esteban:2020cvm, NuFit}.
However, the constraints on \dCP{} are still limited and different measurements are not fully consistent, yet. 
In addition, a degeneracy on \dCP{} stems from the still unknown neutrino mass ordering, leading to two \dCP{} estimates from each measurement~\cite{T2K:2023smv, NOvA:2021nfi, Super-Kamiokande:2017yvm, deSalas:2020pgw, Esteban:2020cvm, NuFit}.
We refer to the case of $\Delta m^2_{31} > 0$ as Normal Ordering (NO) and to $\Delta m^2_{31} < 0$ as Inverted Ordering (IO), with $\Delta m^2_{31}=m^2_3 - m^2_1$ and the neutrino eigenstate masses $m_i$.\\
Liquid scintillator detectors of a few kilotons are sensitive to the low-energy end of the atmospheric neutrino flux~\cite{JUNO:2021tll, Ivanova:2024quh}.
Below $10\,\mathrm{GeV}$, the flux is a mixture of electron and muon neutrinos and the corresponding antineutrinos (\nue, \nueb, \numu, \numub)~\cite{Honda2015, Honda2011, Honda2007, Honda2004}. 
The flavour ratio is about $\frac{\numu + \numub}{\nue+\nueb} \approx 2$, driven by the production mechanism.
Neutrino detectors observe atmospheric neutrinos coming from all directions~\cite{JUNO:2021tll, JUNO:2021vlw, JUNO:2015zny}, as neutrino absorption in the Earth is negligible.
However, flavour conversions~\cite{Wolfenstein:1977ue, Mikheyev:1985zog, Mikheyev1986} are introduced through matter oscillations for neutrinos passing near to the Earth's core, while those are small for  
neutrinos penetrating only the atmosphere or Earth's crust.
The former are referred to as up-going neutrinos, while the latter are referred to as down-going neutrinos, refering to the track direction in the observatory's local coordinate system. 
The flavour conversion probabilities depend on the neutrino energy and the neutrino's path through the Earth~\cite{Barger:1980tf}, and hence on the neutrino direction.
The phase \dCP{} changes the oscilltaion and conversion probabilities, altering the flavour composition arriving at the detector.
Hence, an experiment needs to measure the energy, direction, and flavour of the atmospheric neutrinos in charged current reactions to get a handle on the \dCP{} phase.
Our analysis focuses on the impact of the flavour identification capabilities of the experiments.
We will first estimate the expected number of atmospheric neutrino events in a typical liquid scintillator neutrino detector for different experimental sites.
Next, we consider the impact of the detector's response, e.g., energy and directional resolution, on the measured signal and include background in the analysis.
Then, we model the flavour identification performance.
Finally, we use a Poissonian likelihood to study the sensitivity to \dCP{} under these assumptions for different flavour-identifying performances.

\section{Atmospheric Neutrino Signal}
We use the model of Honda~et~al.~\cite{Honda2015} to estimate the unoscillated (anti-)neutrino fluxes $F_{\nua{\alpha}}(E_{\nu}, \, \theta)$ with $\alpha \in (e, \, \mu)$ as a function of the neutrino energy $E_{\nu}$, and the zenith angle~$\theta$.
The authors~\cite{HondaPage} provide location-specific estimates for different detector sites.
We incorporate the flavour conversion probabilities $\overset{\brabar}{p}_{\beta \rightarrow \alpha}(E_{\nu}, \, \theta | \, \dCP)$ with $\beta \in (e, \, \mu)$, calculated numerically using \emph{nuCraft}~\cite{WALLRAFF2015185}.
%The calculation assumes a PREM Earth density profile~\cite{DZIEWONSKI1981297} and the PMNS parameters from the \emph{NuFIT~5.2}~\cite{NuFit} results\footnote{Including the Super-K atmospheric data}, while we vary \dCP{} between $-180^{\circ}$ and~$180^{\circ}$. %in $10^{\circ}$ steps. 
The calculation assumes a PREM Earth density profile~\cite{DZIEWONSKI1981297} and the PMNS parameters from \emph{NuFIT~5.2}~\cite{NuFit}, while we vary \dCP{} between $-180^{\circ}$ and~$180^{\circ}$. %in $10^{\circ}$ steps. 
For each mass ordering, we consider both NuFIT 5.2 results, including the Super-K atmospheric data and the one excluding it. 
The former gives a $\theta_{23}$ in the lower octant, while the latter yields a $\theta_{23}$ in the upper octant for normal ordering.

\begin{figure}[htbp]
\centering
\includegraphics[width=.99\textwidth]{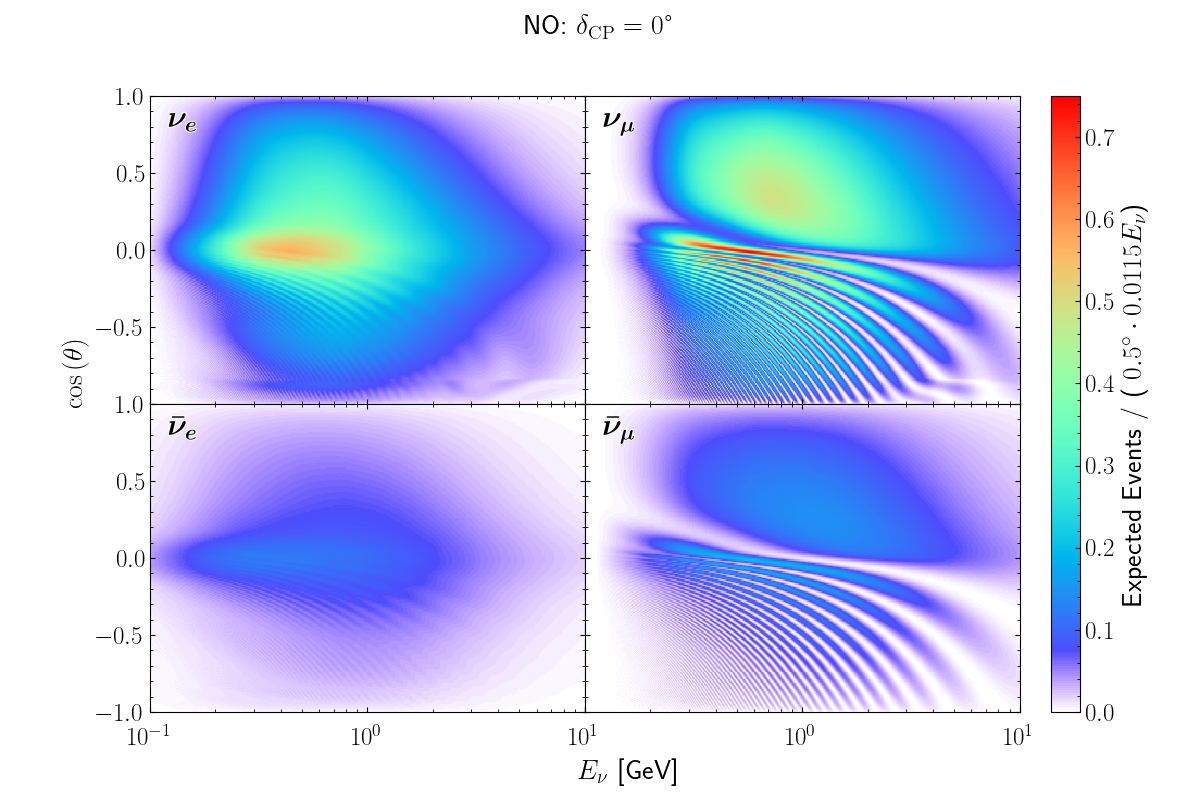}
\caption{Expected number of charged current atmospheric neutrino events versus neutrino energy and zenith angle, assuming an exposure of $250\,\mathrm{kt}\,\mathrm{years}$, for NO, upper octant $\theta_{23}$, and $\dCP{}=0^{\circ}$ for the \emph{SNOLAB}~\cite{Duncan:2010zz} site. 
\emph{Left} electron flavour, \emph{right} muon flavour, \emph{top} neutrinos, and \emph{bottom} antineutrinos.\label{fig:theo_expect_signal}}
\end{figure}

\begin{figure}[htbp]
\centering
\includegraphics[width=.99\textwidth]{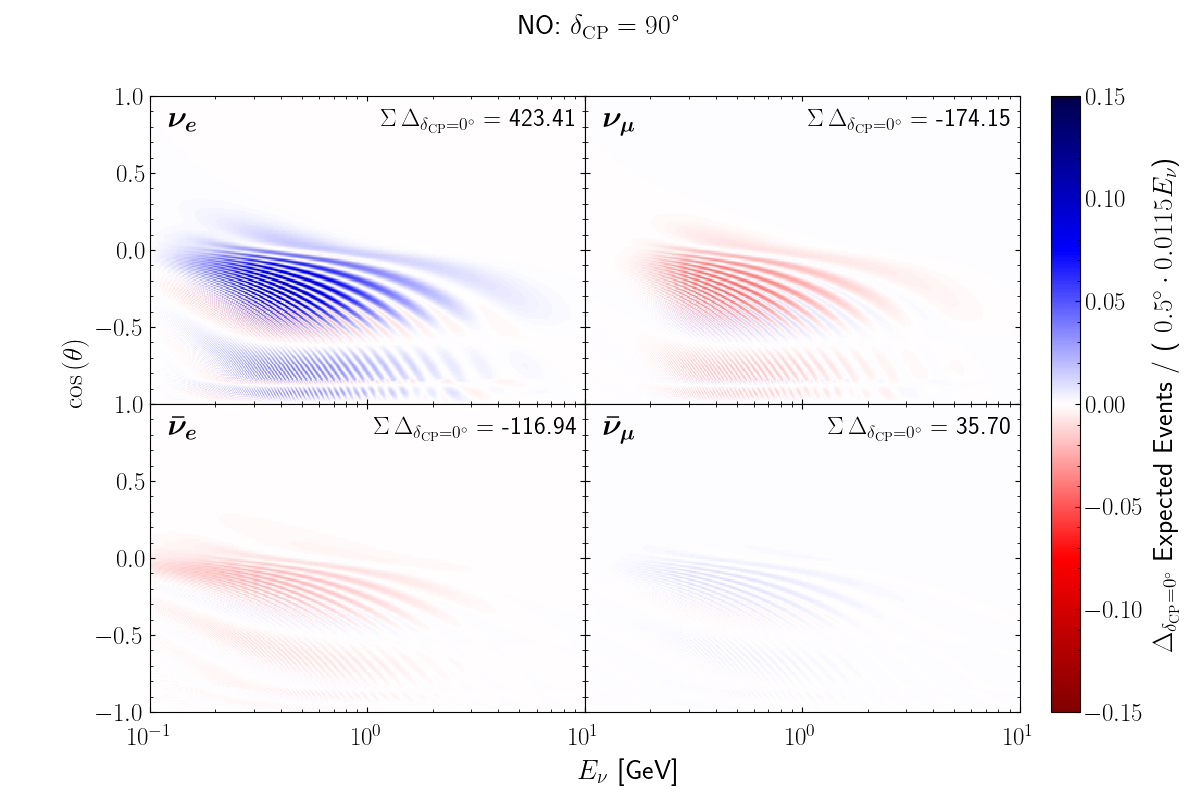}
\includegraphics[width=.99\textwidth]{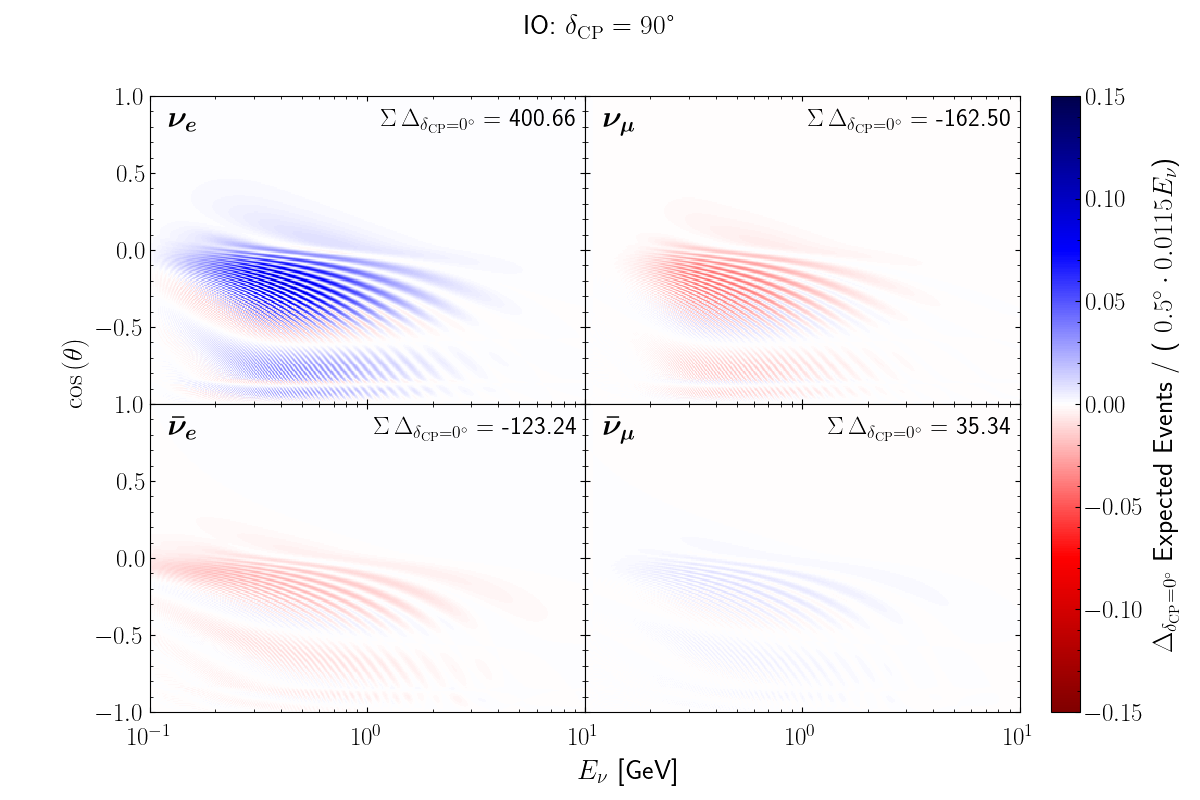}
\caption{Expected difference $\Delta_{\dCP{}=0^{\circ}}$ in the number of charged current atmospheric neutrino events versus neutrino energy and zenith angle for $\dCP{} = 90^{\circ}$ with respect to $0^{\circ}$ and upper octant~$\theta_{23}$ for the \emph{SNOLAB}~\cite{Duncan:2010zz}. An exposure of $250\,\mathrm{kt}\,\mathrm{years}$ is assumed. \emph{Upper plot} NO and \emph{lower plot} IO. For both: \emph{Left} side electron flavour, \emph{right} side muon flavour, \emph{top} neutrinos, and \emph{bottom} antineutrinos. \label{fig:theo_dCP_sig_diff}}
\end{figure}

We calculate the expected number $S^{\nua{\alpha}}$ of charged current atmospheric (anti-)neutrino events in a predefined binning in $E_{\nu}$ and $\theta$, assuming an exposure of $M_{\text{exp}} = 250\,\mathrm{kt}\,\mathrm{years}$ as
\begin{equation}
    S^{\nua{\alpha}}_{ij}(\dCP) = 2\pi \, M_{\text{exp}} \int\limits_{E_{\nu, \, i}}^{E_{\nu, \, i+1}} \int\limits_{\theta_j}^{\theta_{j+1}} dE_{\nu} \, d\theta \, \sigma_{\mathrm{CC}, \, \mathrm{eff}}^{\nua{\alpha}}(E_{\nu}) \sum\limits_{\beta} \overset{\brabar}{p}_{\beta \rightarrow \alpha}(E_{\nu}, \, \theta | \, \dCP) F_{\nua{\beta}}(E_{\nu}, \, \theta), 
\end{equation}
with the effective charged current interaction cross section  $\sigma_{\mathrm{CC}, \, \mathrm{eff}}^{\nua{\alpha}}$, and bins in energy and zenith angle defined by $E_{\nu, \, i}$ and $\theta_j$.
All events with an interaction point in the detector are included, independent of the containment of the whole event.
We take the cross sections from the GENIE v3.4 \emph{G18\_02a\_00\_000} model tuning~\cite{GENIE2021}. 
The effective cross section $\sigma_{\mathrm{CC}, \, \mathrm{eff}}^{\nua{\alpha}}$ per target mass is calculated for a typical organic liquid scintillator mixture with $12\,\%$ hydrogen and $88\,\%$ carbon mass fractions.
As an example, the results for NO and upper octant $\theta_{23}$ with $\dCP{}=0^{\circ}$ are shown in Figure~\ref{fig:theo_expect_signal} for the \emph{SNOLAB}~\cite{Duncan:2010zz}.
The impact of \dCP{} on the expected number of charged current events is shown in Figure~\ref{fig:theo_dCP_sig_diff}.
The CP violation manifests itself in a difference between the flavour oscillations in neutrinos and antineutrinos ($p_{\alpha \rightarrow \beta} \neq \bar{p}_{\alpha \rightarrow \beta}$).
Varying \dCP{} leads to anti-correlated changes in the up-going electron and muon flavour events, while antineutrino are effected in the opposite direction.
For a given neutrino flavour, the changes in the event rate can be positive as well as negative depending on $E_{\nu}$ and $\theta$, leading to wave-like structures in the $E_{\nu}, \, \theta$ space on top of the nominal (e.g., $\dCP = 0^{\circ})$ oscillation pattern.
The total change in the number of (up-going) events is indicated in Figure~\ref{fig:theo_dCP_sig_diff} for each flavour.

\section{Detector Resolution}
\begin{figure}[htbp]
\centering
\includegraphics[width=.7\textwidth]{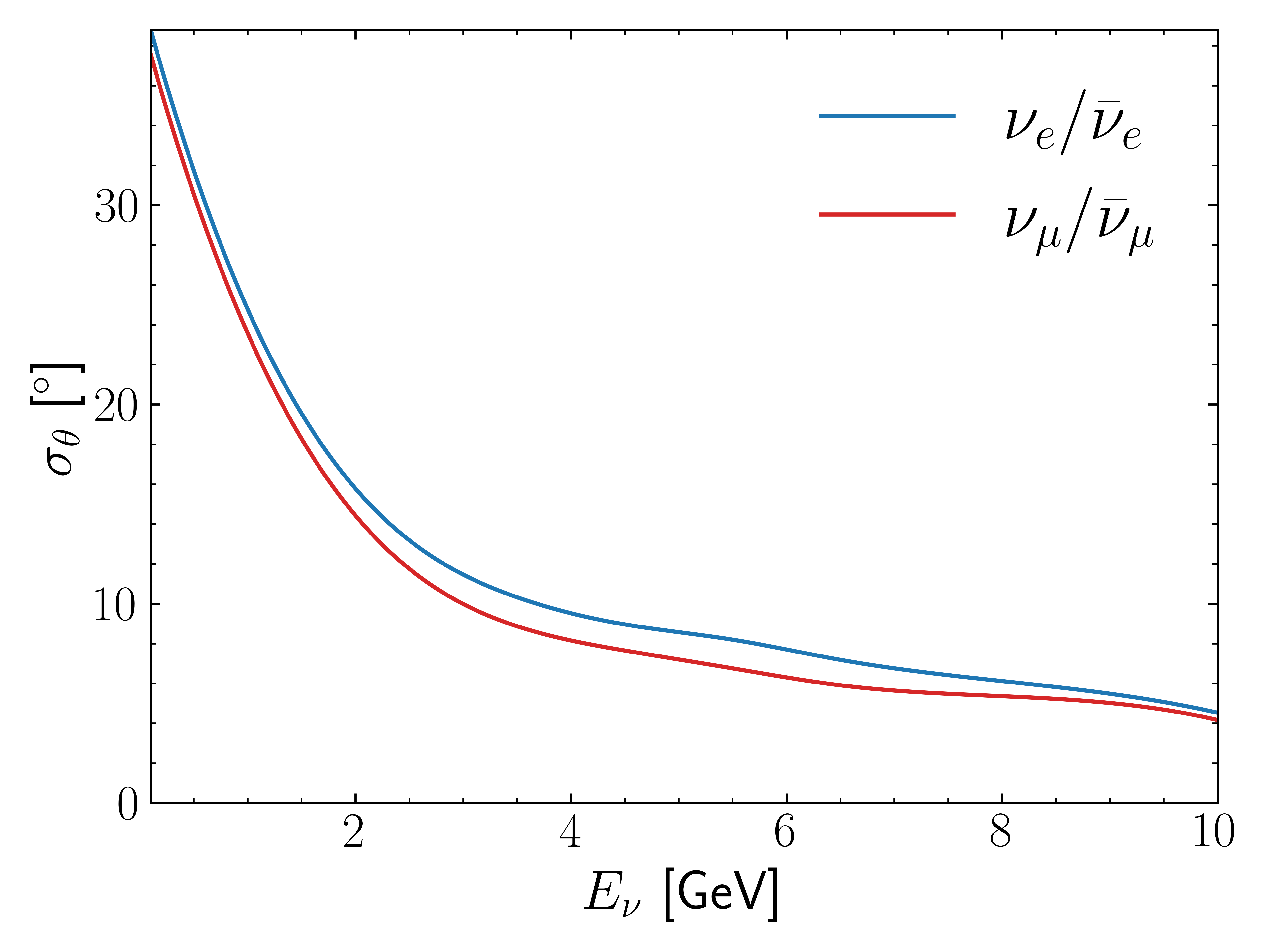}
\caption{Zenith angle resolution versus neutrino energy. 
Given by extrapolating the resolution shown in Figure~4 of \cite{JUNOzenith} using a cubic spline.\label{fig:zenith_res}}
\end{figure}

\begin{figure}[htbp]
\centering
\includegraphics[width=.99\textwidth]{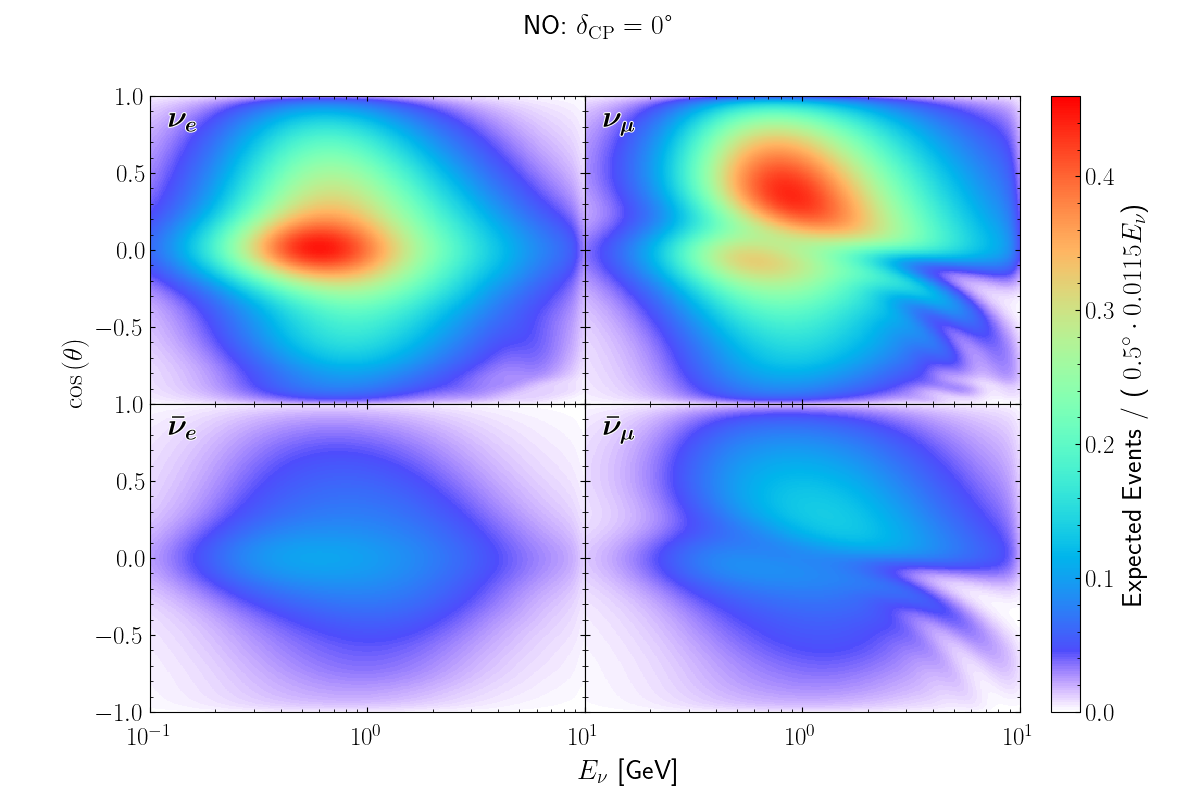}
\includegraphics[width=.99\textwidth]{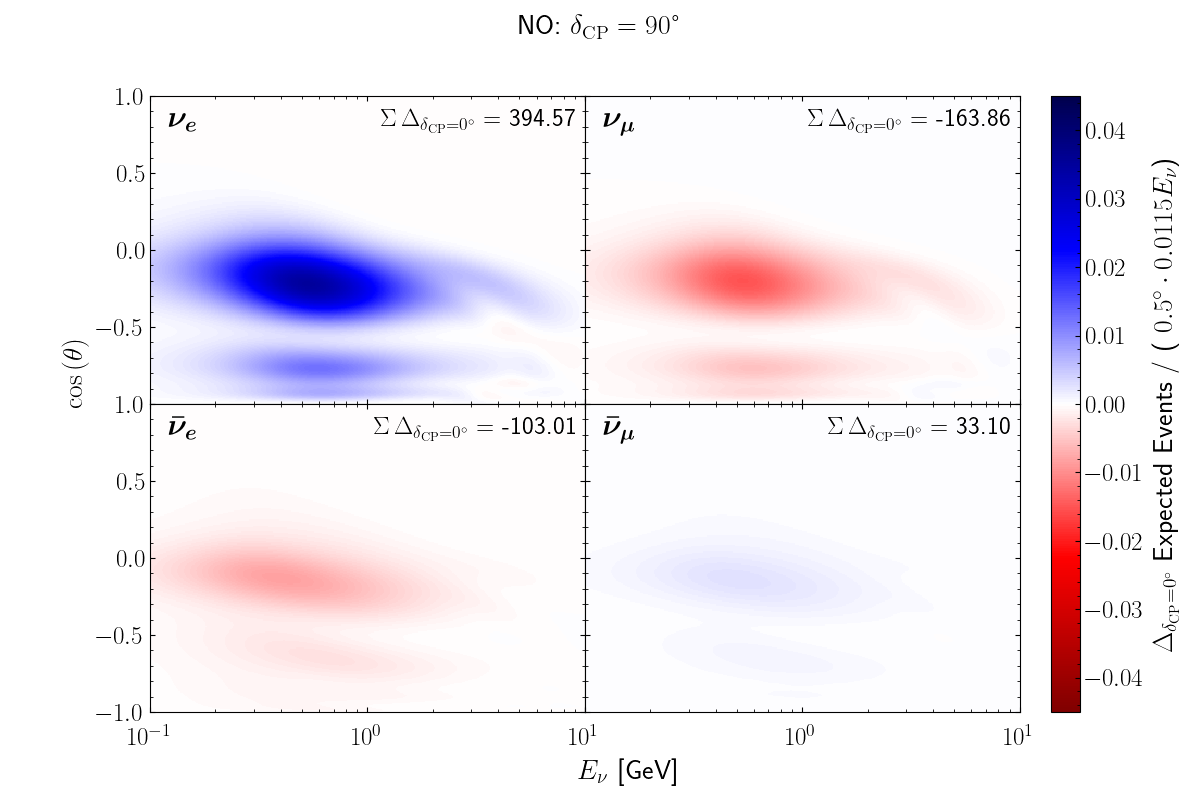}
\caption{\emph{Upper plot:} Expected number of charged current atmospheric neutrino events versus neutrino energy and zenith angle with realistic detector resolution,  NO, upper octant $\theta_{23}$, and $\dCP{}=0^{\circ}$ for the \emph{SNOLAB}~\cite{Duncan:2010zz} site. \emph{Lower plot:} Difference $\Delta_{\dCP{}=0^{\circ}}$ in the number of events for $\dCP{} = 90^{\circ}$ with respect to $0^{\circ}$. An exposure of $250\,\mathrm{kt}\,\mathrm{years}$ is assumed. \emph{Left} electron flavour, \emph{right} muon flavour, \emph{top} neutrinos, and \emph{bottom} antineutrinos.\label{fig:smeared_expec}}
\end{figure}

Any neutrino detector measures the neutrino energy and zenith angle with some resolution $\sigma_{E_{\nu}}$ and $\sigma_{\theta}$.
The resolution for an individual event is a complex function of the event's topology.
Even, different reconstruction algorithms might yield different resolutions, albeit using the same data.
For the study presented here, we use an approximate resolution $\sigma_{E_{\nu}}(E_{\nu})$,  $\sigma_{\theta}(E_{\nu})$ that only depends on the neutrino energy.
We extrapolate the JUNO zenith angle resolution presented by D.~Hongyue~et~al.~\cite{JUNOzenith}, as shown in Figure~\ref{fig:zenith_res}.
%\begin{figure}[htbp]
%\centering
%\includegraphics[width=.99\textwidth]{plots/zenith_res.png}
%\caption{JUNO's zenith resolution as a function of the neutrino energy. \emph{Left}: muon \mbox{(anti-)neutrinos}. \emph{Right}: electron (anti-)neutrinos. The figure is a copy of Figure~4 in~\cite{JUNOzenith}.\label{fig:zenith_res}}
%\end{figure}
Below $10 \, \mathrm{GeV}$ the direction resolution is driven by the angle between the neutrino and the final state lepton~\cite{Super-Kamiokande:1998tou}.
We take the average of the three algorithms presented and interpolate with a cubic spline.
We use the same approach to extrapolate the zenith angle resolution to energies below $1\,\mathrm{GeV}$.
For the energy resolution, we assume a relative uncertainty of $10\,\%$, $\sigma_{E_{\nu}}(E_{\nu}) = 0.1\, E_{\nu}$, motivated by Figure~7-6 in~\cite{JUNO:2015zny}.
We fold the distribution of expected events with a two-dimensional normal distribution corresponding to those resolutions.
The resulting distributions are shown in Figure~\ref{fig:smeared_expec}.
%The corresponding difference to the $\dCP{} = 90^{\circ}$ case is shown below.
The fine structures of the oscillations are smeared out. 
However, multiple regions in the $E_{\nu}, \, \theta$ space with either a clear deficit or excess of expected events remain. 

\section{Backgrounds}
We expect the background to be dominated by neutral current interaction of the same atmospheric neutrinos.
The neutral current interactions form high-energetic hardronic cascades.
These are missing the lepton, determining the neutrino flavour.
However, the cascades can mimic charged current events through tracks from high-energy pions~\cite{JUNO:2015zny}.
Reliable estimates of the background rates are challenging.
We calculate the expected number of neutral current events from the neutral current cross section and flux estimates.
The background estimate depends strongly on the selection algorithm and the hadronization model.
These models typically come with large uncertainties~\cite{GENIE2021}.

\begin{figure}[htbp]
\centering 
\includegraphics[width=.99\textwidth]{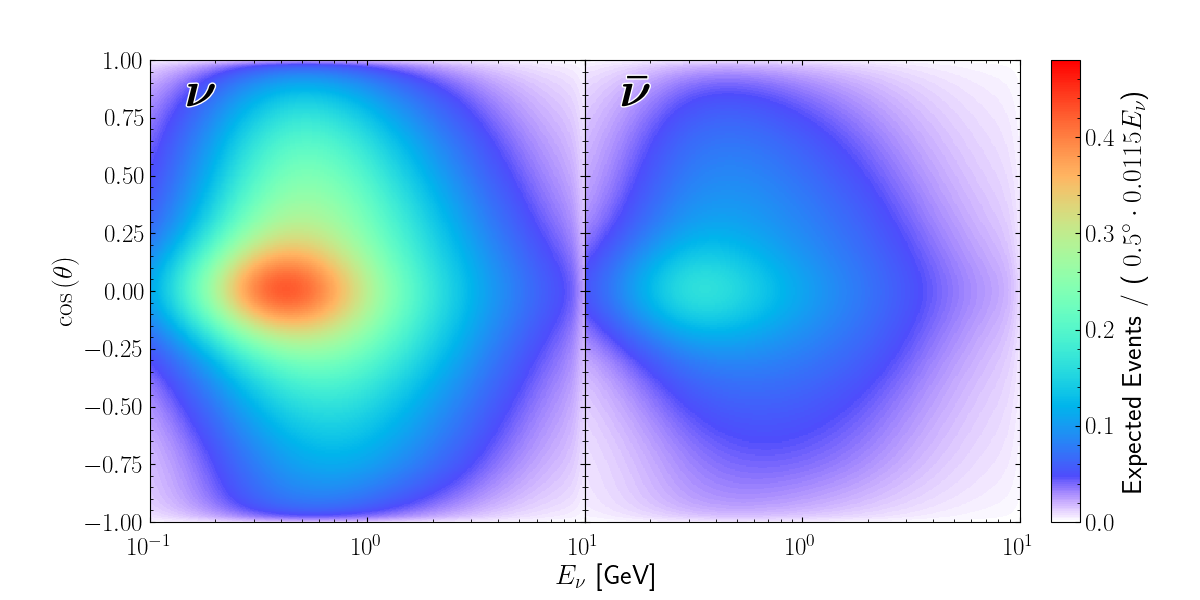}
\caption{Expected number of neutral current atmospheric neutrino events versus neutrino energy and zenith angle, assuming an exposure of $250\,\mathrm{kt}\,\mathrm{years}$ for the \emph{SNOLAB}~\cite{Duncan:2010zz} site. \emph{Left} side neutrino, and \emph{right} side antineutrino. A fraction of these events can mimic charged current events.\label{fig:bkg_expec}}
\end{figure}

Figure~\ref{fig:bkg_expec} shows the distribution $B^{\nua{}}$ of the neutral current events, estimated as
\begin{equation}
    B^{\nua{}}_{ij} = 2\pi \, M_{\text{exp}} \int\limits_{E_{\nu, \, i}}^{E_{\nu, \, i+1}} \int\limits_{\theta_j}^{\theta_{j+1}} dE_{\nu} \, d\theta \, \sigma_{\mathrm{NC}, \, \mathrm{eff}}^{\nua{}}(E_{\nu}) \sum\limits_{\alpha} F_{\nua{\alpha}}(E_{\nu}, \, \theta), 
\end{equation}
with $\sigma_{\mathrm{NC}, \, \mathrm{eff}}^{\nua{}}(E_{\nu})$ the neutral current cross section per target mass, which we also take from  \mbox{GENIE v3.4} \emph{G18\_02a\_00\_000} model tuning~\cite{GENIE2021}.
We apply the same detector response model as for the charged current interactions.
We assume that $5 \, \%$ of the neutral current events will be misidentified into each signal class and that the signal selection efficiency will be $80\,\%$.

\section{CP-violation Sensitivity}
%We perform a likelihood ratio test to determine how significantly a given \dCP{} value deviates from the CP-conserving cases of  $\dCP{} \in \{0^{\circ}, \: 180^{\circ}\}$.
We perform a likelihood ratio test to determine how significantly a given \dCP{} value deviates from the CP-conserving cases of  $\dCP{} = 0^{\circ}$ or $\dCP{} = 180^{\circ}$.
Events selected in each neutrino class~(\nue,~\nueb,~\numu,~\numub) will be a mixture of the anticipated neutrino flavour, misidentified charged current events of other flavours, and background events.
The expected number of events per bin for each class is
\begin{equation} \label{eq:expec}
    \tilde{S}^{\nua{\alpha}}_{ij}(\dCP; \, N_F) = N_F \left[ \epsilon_S \left( 
    \sum\limits_{\nu \: \mathrm{and} \: \bar{\nu}, \beta} \mathcal{E}_{\alpha \beta} S^{\nua{\beta}}_{ij}(\dCP)
    \right) + r_b \left( B^{\nu}_{ij} + B^{\bar{\nu}} _{ij}\right) \right],
\end{equation}
with the signal selection efficiency $\epsilon_s = 0.8$, the background selection rate $r_b = 0.05$, and the flux normalization $N_F$.
$\mathcal{E}_{\alpha \beta}$ accounts for the flavour identification performance with
\begin{equation}
    \mathcal{E}_{\alpha \beta} =
  \begin{cases}
    \alpha_f \, \alpha_{\pm}, & \text{matching}\: \nu \: \text{or} \: \bar{\nu}, \: \beta = \alpha \\ 
    \alpha_f \, (1-\alpha_{\pm}), & \text{not matching}\: \nu \: \text{or} \: \bar{\nu}, \: \beta = \alpha \\
    (1-\alpha_f) \, \alpha_{\pm}, & \text{matching}\: \nu \: \text{or} \: \bar{\nu}, \: \beta \neq \alpha \\
    (1-\alpha_f) \, (1-\alpha_{\pm}), & \text{not matching}\: \nu \: \text{or} \: \bar{\nu}, \: \beta \neq \alpha \\
  \end{cases},
\end{equation}
and the flavour identification accuracy $\alpha_f$ and the neutrino-antineutrino identification accuracy $\alpha_{\pm}$.
For an ideal detector we would have {$\alpha_f=\alpha_{\pm}=1$}.

\begin{figure}[htbp]
\centering
\includegraphics[width=.99\textwidth]{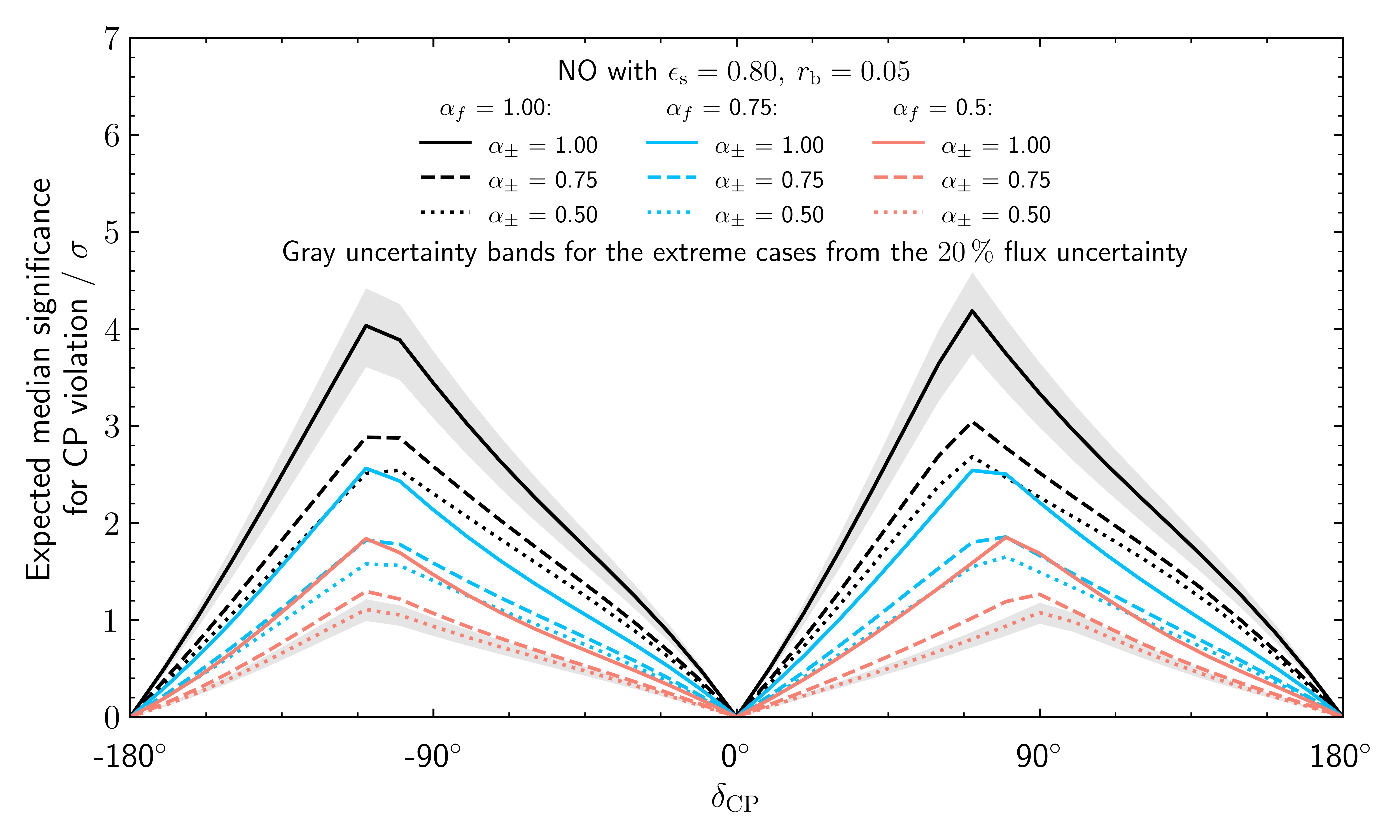} \\
\includegraphics[width=.99\textwidth]{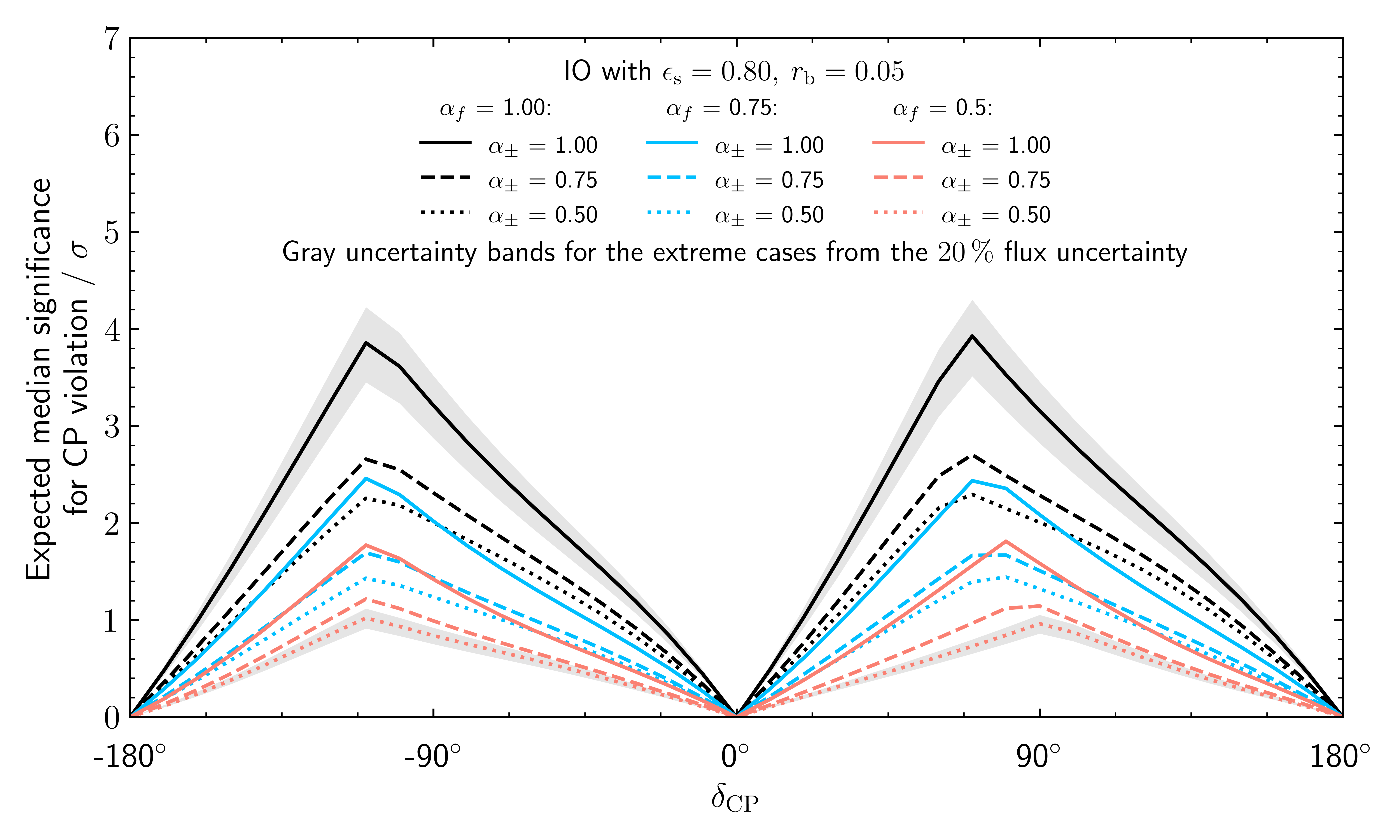}
\caption{CP violation sensitivity. \emph{Top} NO and \emph{bottom} IO. We assume our baseline background model with $\theta_{23}$ in the upper octant and a detector located at the \emph{SNOLAB}~\cite{Duncan:2010zz} site and vary the flavour identification accuracies. The uncertainty band is given by the $\pm 20 \, \%$ flux uncertainty.\label{fig:dCP_sens}}
\end{figure}

For our nominal flux model, $N_F = 1$.
Varying $N_F$ allows to model the average flux uncertainty as well as the average cross section uncertainty.
We assume an common uncertainty of $20\,\%$ for $N_F$~\cite{Honda2015} and treat $N_F$ as a nuisance parameter.
The likelihood ratio between the true $\delta_{\mathrm{CP}}^0$~hypothesis and an alternative hypothesis~$\delta_{\mathrm{CP}}^{\mathrm{test}}$, with the expectation values for the true hypothesis as the reference sample (Asimov data) is
\begin{align}
    \begin{split} \label{eq:llhratio}
    q(\dCPt |\, \dCPn) &= -2 \log \frac{\sup\limits_{N_F} \mathcal{L}(\dCPt; \, N_F)}{\mathcal{L}(\dCPn; \, N_F^0)}  \\ 
      &= -2 \sup\limits_{N_F} \sum_{\nu \: \text{and} \: \bar{\nu}, \atop \alpha, \, i, \, j} \left[ \tilde{S}^{\nua{\alpha}}_{ij}(\dCPn; \, N_F^0)  \log \left( \frac{\tilde{S}^{\nua{\alpha}}_{ij}(\dCPt; \, N_F)}{\tilde{S}^{\nua{\alpha}}_{ij}(\dCPn; \, N_F^0)} \right) \right.  \\ 
       &{} \hspace{92pt} + \left. \tilde{S}^{\nua{\alpha}}_{ij}(\dCPn; \, N_F^0) - \tilde{S}^{\nua{\alpha}}_{ij}(\dCPt; \, N_F)  \right],
    \end{split}
    %Etrue*np.log(Etest/Etrue) + Etrue-Etest).sum()
\end{align}
given by the Poissonian fluctuation for each bin, with the injected flux normalization of the true hypothesis $N_F^0$.
The median expected significance of excluding $\dCPt{}$ when $\dCPn{}$ is true is, therefore, $\sqrt{q(\dCPt |\, \dCPn)}$, given by Wilks' Theorem~\cite{Wilks:1938dza}. 
The significance for CP violation is thus
\begin{equation}
\min \left\{ \sqrt{q(\dCPt |\, 0^{\circ})}, \sqrt{q(\dCPt |\, 180^{\circ})} \right\}.
\end{equation}
Figure~\ref{fig:dCP_sens} shows the median expected sensitivity as a function of \dCP{} for NO and IO and a set of different flavour identification performances. 
The maximal sensitivities, which are in proximity to the two maximal CP-violating cases of $\dCP{} \in \left\{ -90^{\circ}, \: 90^{\circ}\right\}$, vary between $4 \, \sigma$ in the best case and $1 \, \sigma$ in the worst case.
Decreasing flavour and neutrino-antineutrino identification performance decreases the CP violation sensitivity as expected.
In general, the sensitivity is smaller in the IO case.

\begin{figure}[htbp]
\centering
\includegraphics[width=.99\textwidth]{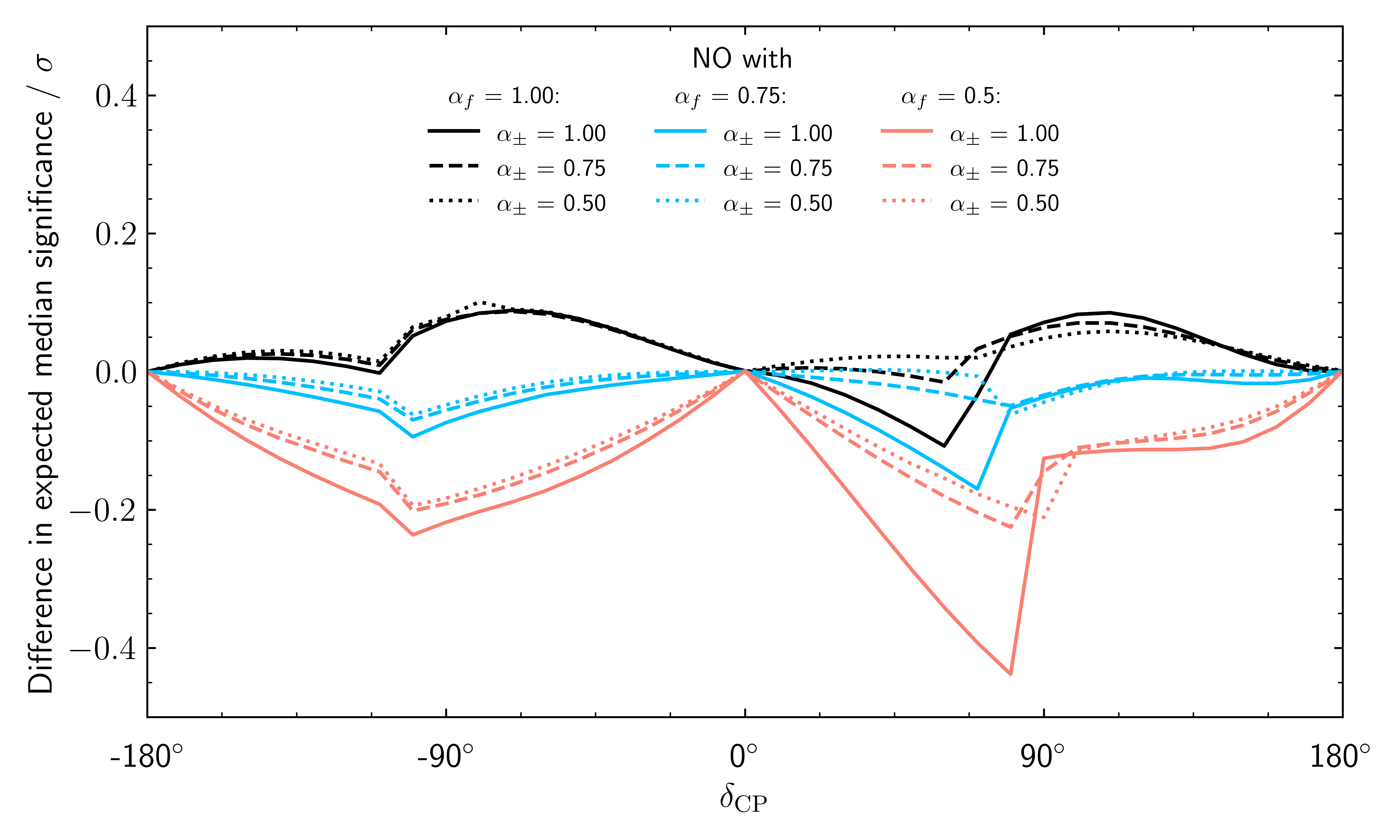} 
\caption{Sensitivity alteration from injecting \emph{NuWro} into the Asimov data. We assume our background model with $\theta_{23}$ in the upper octant and a detector located at the \emph{SNOLAB}~\cite{Duncan:2010zz} site, and vary the flavour identification accuracies. \label{fig:xsec_impact}}
\end{figure}

We study the impact of the cross section uncertainties. $\tilde{S}^{\nua{\alpha}, \: NuWro}_{ij}(\dCP; \, N_F)$ is the predicted number of events using the \emph{NuWro} \cite{Golan:2012rfa} cross section \cite{Golan:2012wx} for the neutrino charged current interaction, estimated as \ref{eq:expec}.
Further, $\tilde{S}^{\nua{\alpha}, \: Nuance}_{ij}(\dCP; \, N_F)$ is the predicted number of events using the \emph{Nuance} \cite{Casper:2002sd} cross section \cite{Formaggio:2012cpf} for all neutrino interactions.
We extend the likelihood ratio estimation \ref{eq:llhratio} by replacing the \emph{GENIE} prediction $\tilde{S}^{\nua{\alpha}}_{ij}(\dCP; \, N_F)$ with 

\begin{align}
     \begin{split}
     \tilde{S}^{\nua{\alpha}, \: \mathrm{Sys}}_{ij}(\dCP; \, N_F, p_{\mathrm{NuWro}}, p_{\mathrm{Nuance}}) = &\left(1-p_{\mathrm{NuWro}} - p_{\mathrm{Nuance}} \right) \tilde{S}^{\nua{\alpha}}_{ij}(\dCP; \, N_F) \\
     &+ p_{\mathrm{NuWro}} \, \tilde{S}^{\nua{\alpha}, \: NuWro}_{ij}(\dCP; \, N_F) \\ 
     &+ p_{\mathrm{Nuance}} \, \tilde{S}^{\nua{\alpha}, \: Nuance}_{ij}(\dCP; \, N_F),
    \end{split}
\end{align}
adding $p_{\mathrm{NuWro}}$ and $p_{\mathrm{Nuance}}$ as free fitting parameters.
The injected cross section model is recovered in all fits, i.e., $p_{\mathrm{NuWro}} =  p_{\mathrm{Nuance}} = 0$ for the \emph{GENIE} prediction, $p_{\mathrm{NuWro}} = 1,   p_{\mathrm{Nuance}} = 0$ for the \emph{NuWro} prediction, and $p_{\mathrm{NuWro}} = 0,   p_{\mathrm{Nuance}} = 1$ for the \emph{Nuance} prediction. 
The sensitivity impact is shown in Figure~\ref{fig:xsec_impact}, where the \emph{NuWro} prediction is injected into the Asimov data. 
The cross section uncertainty causes a sensitivity decrease of less than $0.5 \, \sigma$.
The following analysis is conducted using \ref{eq:expec} during fitting and the \emph{GENIE} prediction.
However, an additional sensitivity reduction of less than $0.5 \, \sigma$ is assumed.

\begin{figure}[htbp]
\centering
\includegraphics[width=.99\textwidth]{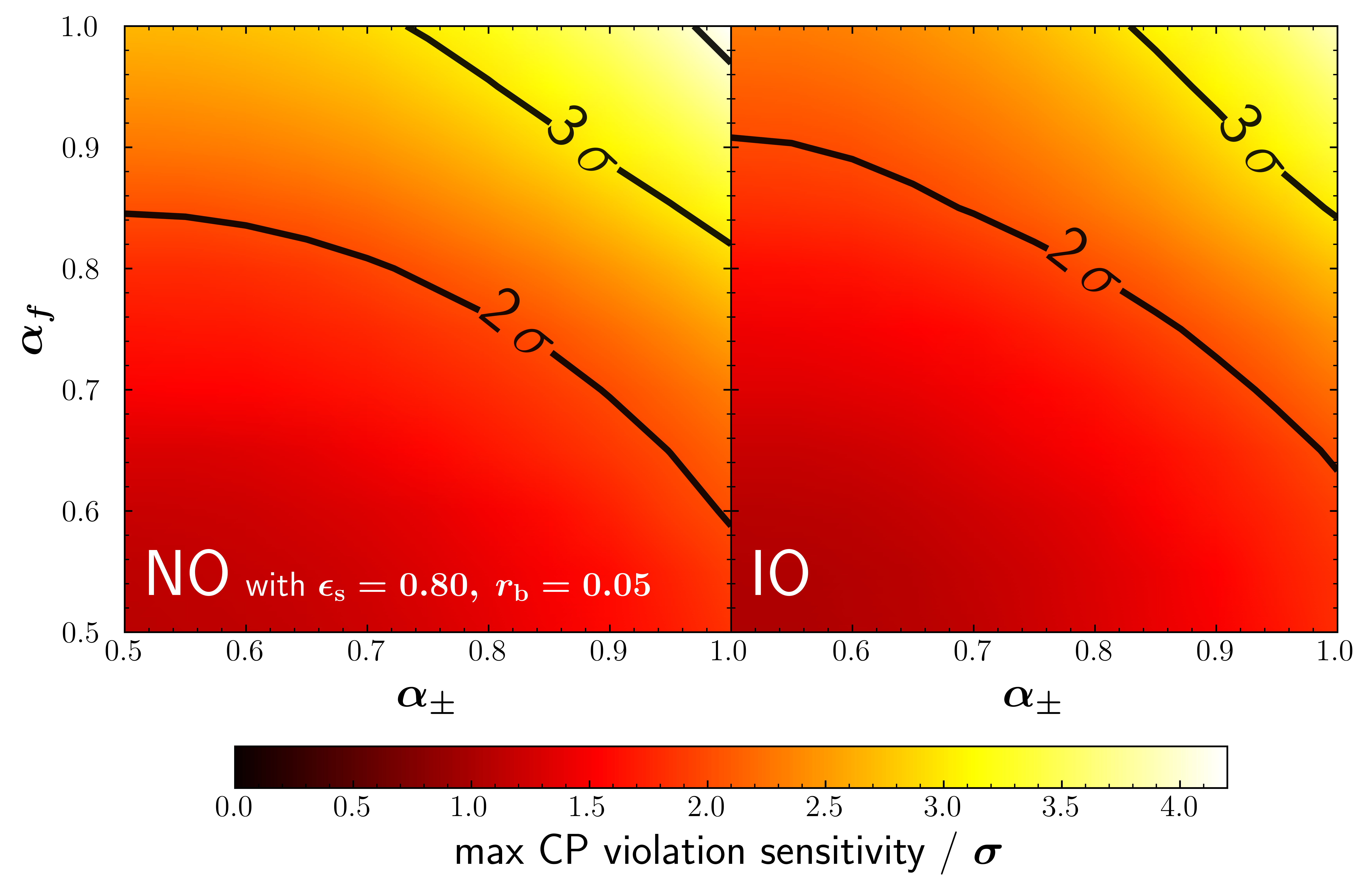}
\caption{Maximal sensitivity to CP violation as a function of $\alpha_{\pm}$ and $\alpha_F$ for the nominal flux model, upper octant of $\theta_{23}$ at the \emph{SNOLAB}~\cite{Duncan:2010zz} site. \emph{Left} NO, and \emph{right} IO.\label{fig:dCP_sens_scan}}
\end{figure}

\begin{figure}[htbp]
\centering
\includegraphics[width=.99\textwidth]{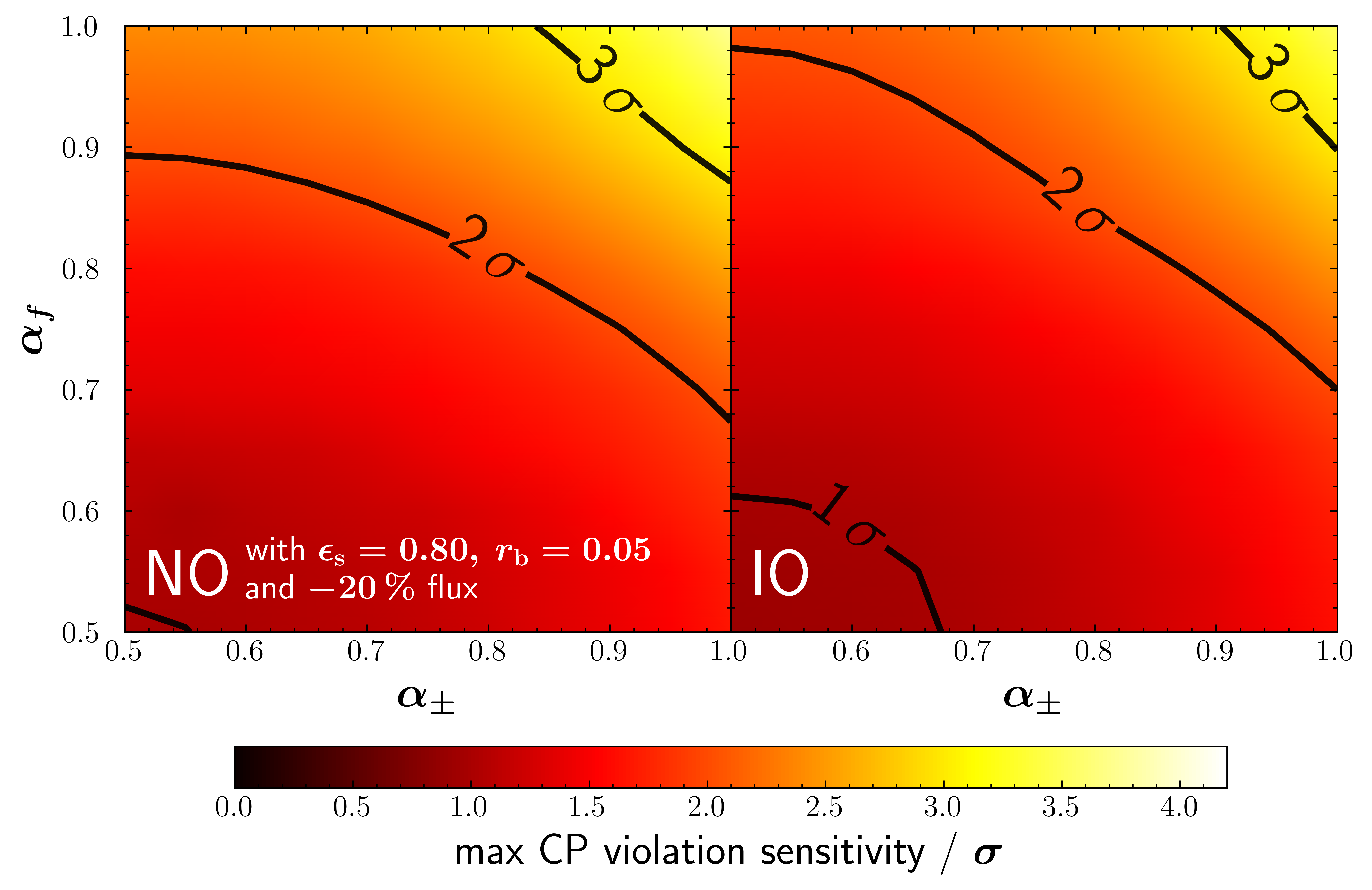} \\
\includegraphics[width=.99\textwidth]{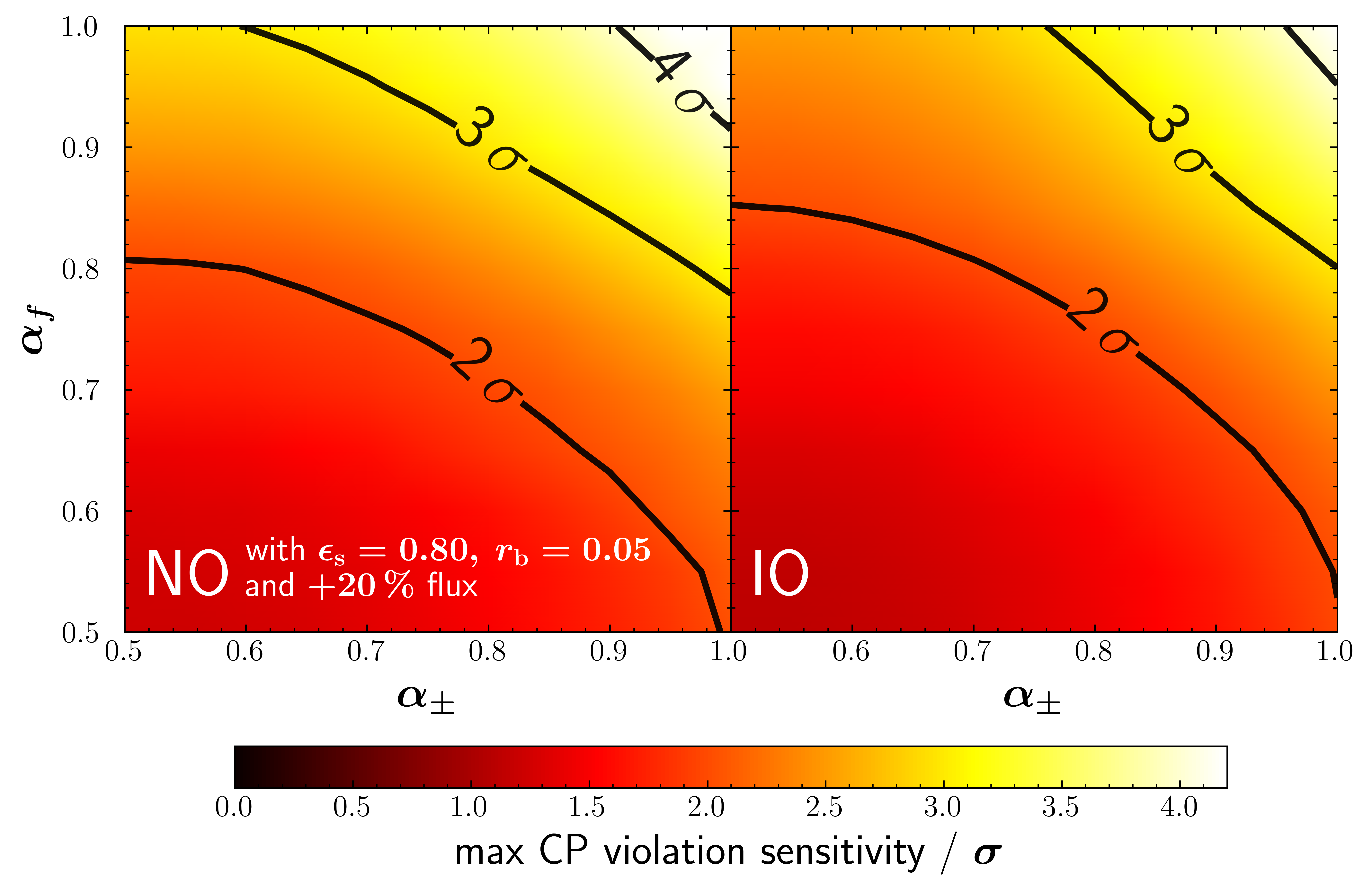} 
\caption{Maximal sensitivity to CP violation as a function of $\alpha_{\pm}$ and $\alpha_F$ for $\pm 20 \, \%$ normalizations of the flux with $\theta_{23}$ in the upper octant at the \emph{SNOLAB}~\cite{Duncan:2010zz} site.. \emph{Left} NO, and \emph{right} IO.\label{fig:dCP_sens_scan_+-flux}}
\end{figure}

\begin{figure}[htbp]
\centering
\includegraphics[width=.99\textwidth]{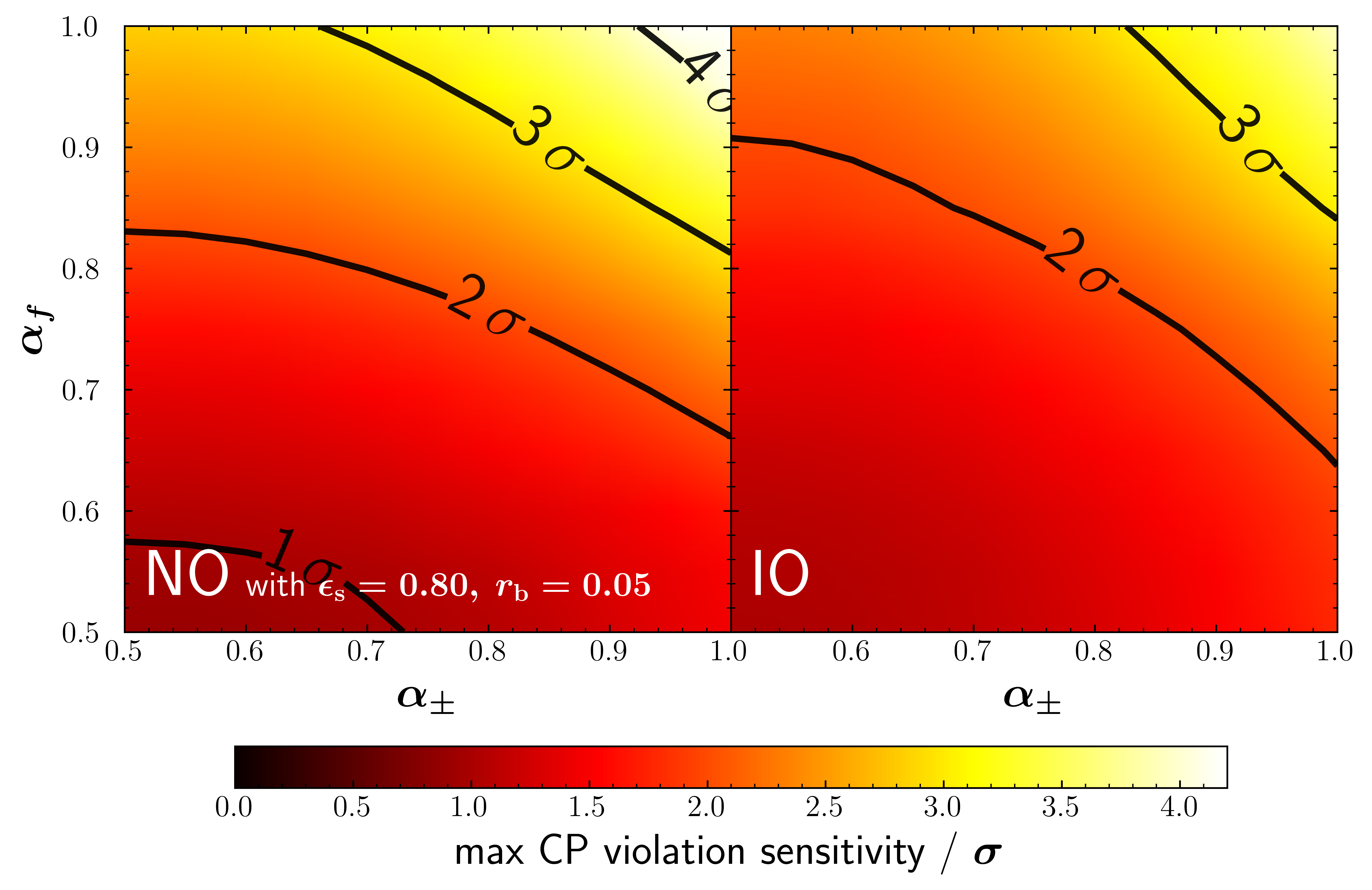}
\caption{Maximal sensitivity to CP violation as a function of $\alpha_{\pm}$ and $\alpha_F$ for the nominal flux model with $\theta_{23}$ in the lower octant for NO at the \emph{SNOLAB}~\cite{Duncan:2010zz} site. \emph{Left} NO, and \emph{right} IO.\label{fig:dCP_sens_scan_SNOLAB_withSKATM}}
\end{figure}

\begin{figure}[htbp]
\centering
\includegraphics[width=.99\textwidth]{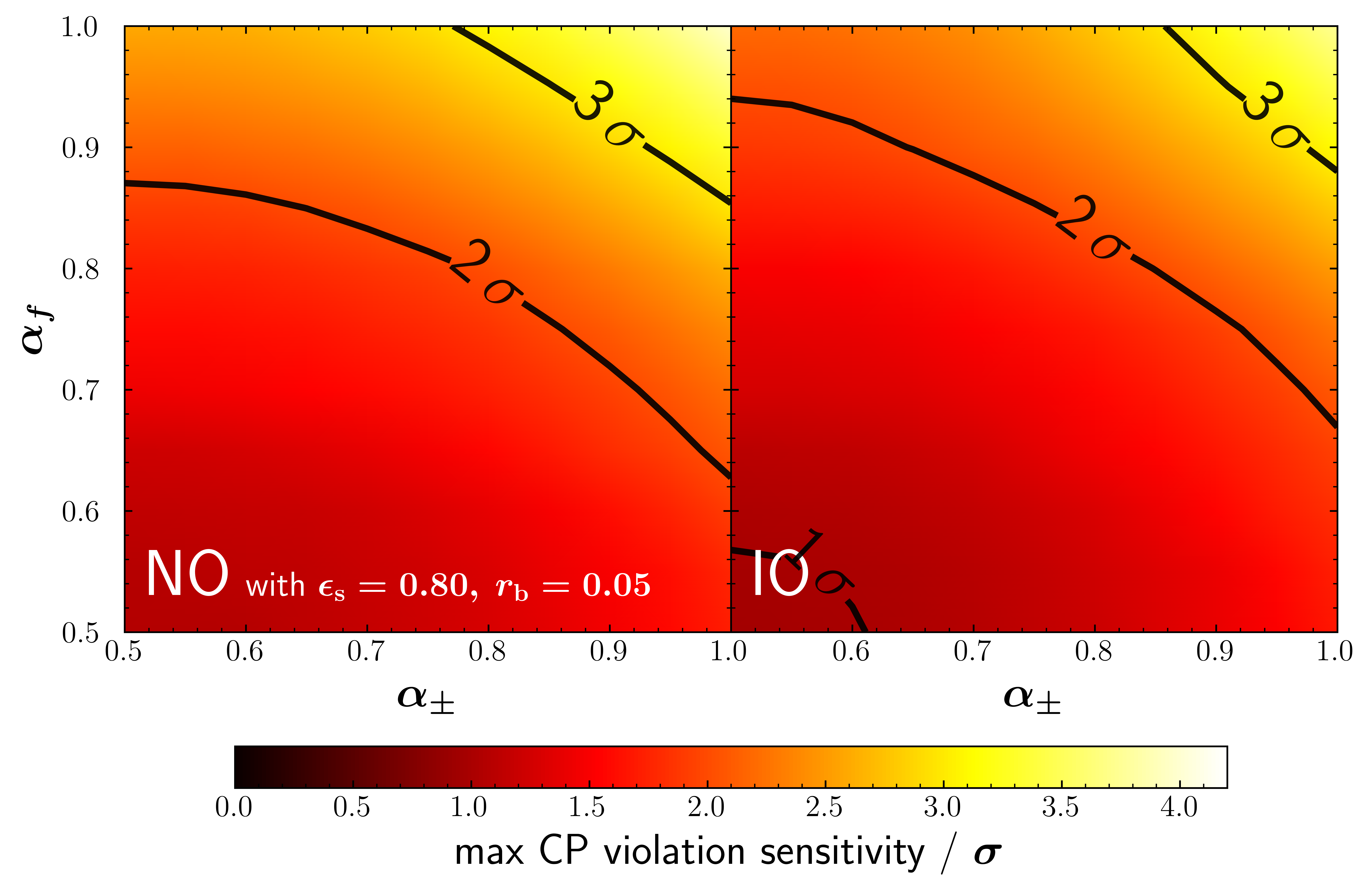} \\
\includegraphics[width=.99\textwidth]{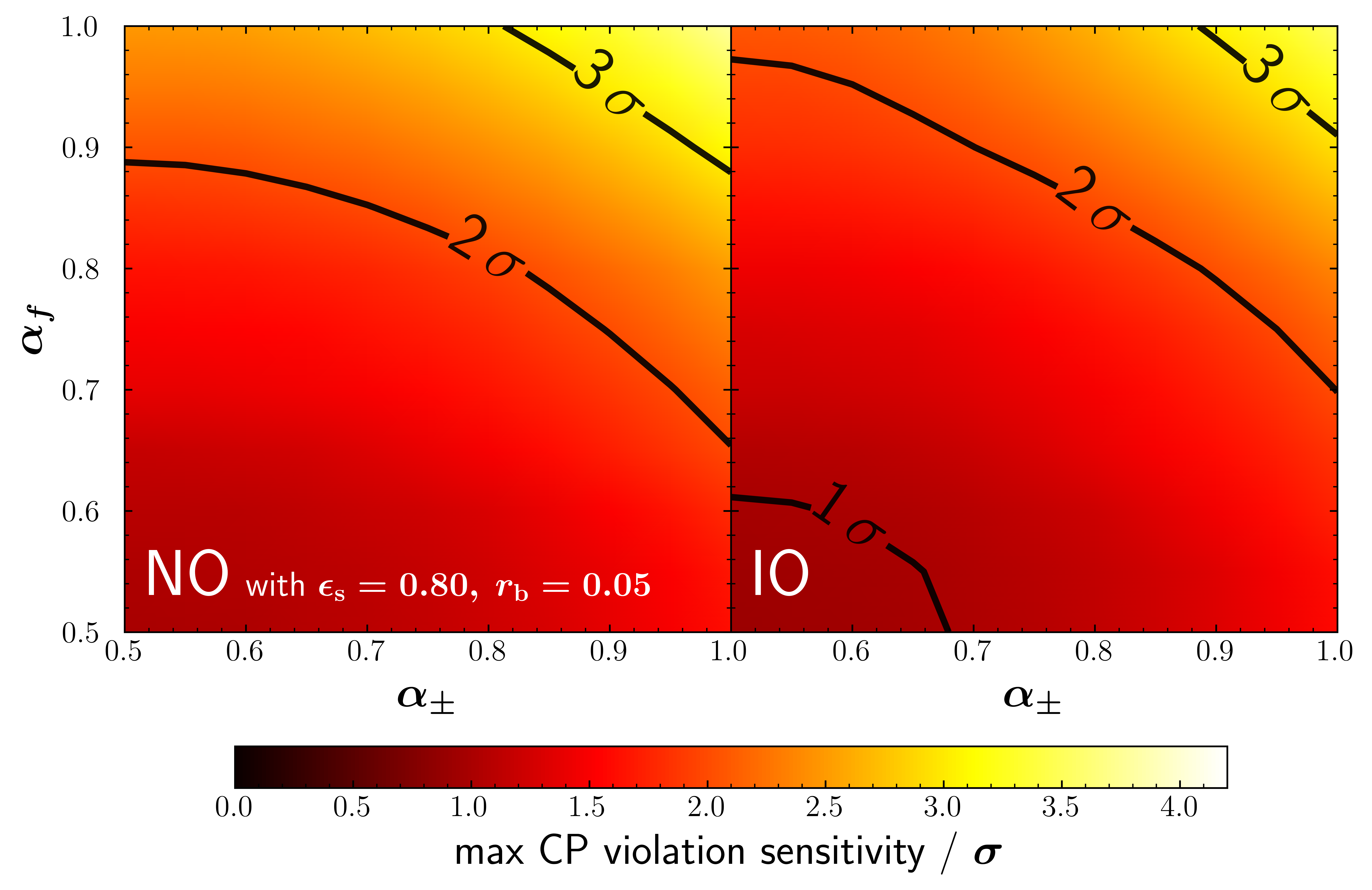} 
\caption{Maximal sensitivity to CP violation as a function of $\alpha_{\pm}$ and $\alpha_f$ for different detector locations with $\theta_{23}$ in the upper octant. \emph{Top} \emph{Gran Sasso National Laboratory}~\cite{gransasso} and \emph{bottom} \emph{Kamioka}~\cite{Kamioka} mine. \emph{Left} NO, and \emph{right} IO.\label{fig:dCP_sens_scan_gransassokamioka}}
\end{figure}

In Figure~\ref{fig:dCP_sens_scan}, we show the maximal CP violation sensitivity as a function of the neutrino flavour identification performance.
An overall identification accuracy of $\alpha_f \approx \alpha_{\pm} \approx 90\, \%$ is required to reach a maximal sensitivity of more than $3 \, \sigma$ with an exposure of $250\,\mathrm{kt}\,\mathrm{years}$.
A significance of $4 \, \sigma$ can only be achieved with accuracies of larger than $95 \, \%$. 
The impact of the flux normalization is shown in Figure~\ref{fig:dCP_sens_scan_+-flux}.
In case of a lower flux, the requirements for the selection performance become more stringent, as expected.
For comparison, Figure~\ref{fig:dCP_sens_scan_SNOLAB_withSKATM} shows the maximal sensitivity to CP violation for the \emph{NuFIT~5.2}~\cite{NuFit} parameters including the Super-K atmospheric neutrinos, i.e., lower octant $\theta_{23}$ for the NO.
The sensitivity slightly increases for NO due to the lower octant $\theta_{23}$.
The IO sensitivity is not altered much as the Super-K data has only little impact on the oscillation parameters for IO, and $\theta_{23}$ is still in the upper octant.
Figure~\ref{fig:dCP_sens_scan_gransassokamioka} shows the maximal sensitivity to CP violation for a detector located at the \emph{Gran Sasso National Laboratory (LNGS)}~\cite{gransasso} and the \emph{Kamioka} mine~\cite{Kamioka}.
The impact of the detector location is minor despite the variation in the initial atmospheric neutrino flux.
%Maximizing the flavour identification performance is a crucial ingredient for the CP violation sensitivity.
%With overall identification accuracies of better than $95 \, \%$, $3 \, \sigma$ sensitivity is reached in all settings.

\section{Summary}
We studied the response of a typical liquid scintillator neutrino detector to the atmospheric neutrino flux with respect to signatures of CP violation. 
We calculated the number of expected charged current events as a function of the neutrino energy and zenith angle 
%for the different neutrino flavours, assuming an exposure of $250\,\mathrm{kt}\,\mathrm{years}$ 
at common neutrino observatory locations.
To account for oscillations of the neutrino in the Earth's matter, we used \emph{nuCraft} with a varying CP violation angle \dCP.
The detector's energy and zenith resolution smears out the fine structure of the \dCP{} signal while cumulated effects remain.
%We constructed a baseline model for the background events originating from neutral current atmospheric neutrino interactions.
We include the background originating from neutral current interactions of atmospheric neutrinos in our analysis.
We apply a Poissonian likelihood ratio test to determine the sensitivity to CP violation.
%This modeling lacks numerous systematic effects of a realistic detector.
Our analysis take only the major systematic effects into account (flux normalization, cross section, octant of $\theta_{23}$).
It should be updated with detector-specific effects.
%Our estimated sensitivities are, hence,  order of magnitude estimates.
%Our estimated sensitivities are, hence, approximations that need to be updated for a specific detector.
The maximal significance for {$\dCP=\pm 90^\circ$} varies between $4 \, \sigma$ and $0.5 \, \sigma$, with an additional reduction of less than $0.5\,\sigma$ given by the cross section uncertainty, depending strongly on the flavour identification capabilities of the detector.
%The maximal sensitivity will be for \dCP{} values close to the maximal CP violation of $\dCP{} \in \left\{ -90^{\circ}, \: 90^{\circ}\right\}$.
The dependence on the neutrino flux and the detector site turned out to be small.
%A slightly higher atmospheric neutrino flux results in a higher sensitivity, while a lower flux decreases the sensitivity.
%However, within the flux normalization uncertainty, the impact is small. The detector location also has a small impact.
An overall identification accuracy of $90\, \%$ is required to gain at least $3 \, \sigma$ sensitivity.

\clearpage

% Bibliography

%% [A] Recommended: using JHEP.bst file
\bibliographystyle{JHEP}
\bibliography{biblio.bib}

%% or
%% [B] Manual formatting (see below)
%% (i) We suggest to always provide author, title and journal data or doi:
%% in short all the informations that clearly identify a document.
%% (ii) please avoid comments such as "For a review'', "For some examples",
%% "and references therein" or move them in the text. In general, please leave only references in the bibliography and move all
%% accessory text in footnotes.
%% (iii) Also, please have only one work for each \bibitem.

%\begin{thebibliography}{99}

%\bibitem{a}
%Author,
%\emph{Title},
%\emph{J. Abbrev.} {\bf vol} (year) pg.

%\bibitem{b}
%Author,
%\emph{Title},
%arxiv:1234.5678.

%\bibitem{c}
%Author,
%\emph{Title},
%Publisher (year).

%\end{thebibliography}
\end{document}